\documentclass[epj]{svjour}
\usepackage{latexsym}
\usepackage{graphics}
\begin{document}
\title{Orbital and spin correlations in
  Ca$_{2-x}$Sr$_x$RuO$_4$: A mean 
  field study}

\author{Manfred Sigrist\inst{1} \and Matthias Troyer\inst{1,2}}
%
%
\institute{Theoretische Physik , ETH-H\"onggerberg
CH-8093 Z\"urich, Switzerland \and 
Computational Laboratory, ETH Z\"urich, CH-8092 Z\"urich, Switzerland}

\date{Received: date / Revised version: date}
%
\abstract{The alloy Ca$_{2-x}$Sr$_x$RuO$_4$ exhibits a complex phase
  diagram with 
peculiar magnetic metallic phases. In this paper some aspects of this
alloy are discussed based on a mean field theory for 
an effective Kugel-Khomskii model of localized orbital and
spin degrees of freedom. This model results from an orbital
selective Mott transition which in the three-band system
localized two orbitals while leaving the third one itinerant. 
Special attention is given to the region around a structure quantum
phase transition 
at $ x \approx 0.5 $ where the crystal lattice changes from tetragonal to
orthorhombic symmetry while leaving the system metallic. This transition
yields, a change from ferromagnetic to antiferromagnetic spin
correlations. The complete 
mean field phase diagram for this transition is given including orbital
and spin order. The anisotropy of spin susceptibility, a consequence
of spin-orbit coupling and orbital correlation, is a tell-tale
sign of one of these phases. In the predominantly
antiferromagnetic phase we describe a metamagnetic transition in a
magnetic field and show that coupling of the itinerant band 
 to the localized degrees of freedom yields an anomalous
longitudinal magnetoresistance transition. Both phenomena are connected with
the evolution  of the ferromagnetic and
antiferromagnetic domains in the external magnetic field and agree
qualitatively with the experimental findings.
}
\maketitle
\section{Introduction}
\label{intro}
\begin{figure}
\resizebox{0.4\textwidth}{!}{%
  \includegraphics{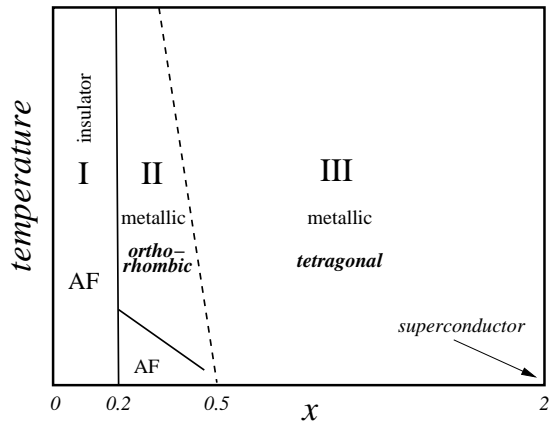}}
\caption{Basic schematic phase diagram of Ca$_{2-x}$Sr$_x$RuO$_4$ at
  low temperatures (following Nakatsuji and Maeno\cite{NAKAMAE}). 
The region III is metallic, paramagnetic and has tetragonal crystal
symmetry. The region II is metallic, at low temperature with
presumably antiferromagnetic correlation and has orthorhombic
symmetry. In both regions the susceptibility over a certain range of
temperature is Curie-like with a spin $S=1/2$. The region I is a Mott
insulator with antiferromagnetic long range order of a spin $S=1$.   
}
\label{NM-PD}
\end{figure}

Layered ruthenate compounds have gained an important position as a 
class of transition metal oxides displaying diverse
physical properties. The interest in these materials has been initiated by 
the discovery of unconventional superconductivity 
in the single-layered Sr$_2$RuO$_4$ which represents most likely a realization
of spin-triplet Cooper pairing, an analog of the A-phase of superfluid
$^3$He.\cite{maeno1,PHYSTODAY} Since the mechanism for pairing is
probably of magnetic origin, the magnetic properties of
related systems have been studied extensively. Increasing the
number of RuO$_2$-layers, following through the Ruddelson-Popper series  
Sr$_{n+1}$Ru$_n$O$_{3n+1}$ with $ n $ as the number of layers per unit
cell, the system turns into a ferromagnet for  $ n \geq 3 $
.\cite{CAO} Peculiar metamagnetic behavior was 
reported for the double-layer system ($n=2$) and was interpreted as a
novel type of quantum critical behavior.\cite{S3R2O7} On the other hand,
replacing Sr by the isoelectronic Ca in the single-layer compound yields a
Mott-insulator with antiferromagnetic long-range order of localized
spin-1 degrees of freedom. 
These contrasting properties of the single-layer compound 
motivated Nakatsuji and Maeno (NM) to investigate 
the continuous series of alloys Ca$_{2-x}$Sr$_x$RuO$_4$.\cite{NAKAMAE}
The interpolating phase diagram is complex. NM 
identified three distinct ranges of $x$ characterized by
different crystal structures as well as electronic and, in particular,
magnetic properties.  
The present study is concerned with a certain range in the NM phase diagram of
these alloys (Fig.\ref{NM-PD}). Therefore it is in order to 
briefly review here the basic experimental results for
Ca$_{2-x}$Sr$_x$RuO$_4$.\cite{NAKAMAE}

Starting from
the metallic and superconducting stoichiometric compound Sr$_2$RuO$_4$ ($x=2$) the
introduction of the smaller ion Ca for Sr leads to a contraction of the
crystal volume. This is achieved by the rotation of the
RuO$_6$-octahedra, which occurs for $ x < 1.5 $.\cite{friedt} 
Nevertheless, up to dopings of $ x = 0.5 $ in Ca$_{2-x}$Sr$_x$RuO$_4$ the
crystal structure remains {\it tetragonal}. This doping range,
called region III by NM, is characterized by an increase of the
uniform susceptibility which eventually becomes
Curie-like with a Curie constant corresponding to nearly free spin $S= 1/2 $
degrees of freedom as $ x=0.5
$ is approached. In the doping range $ 0.2 \leq x < 0.5 $, the region
II following NM,
the system suffers a further crystal deformation through tilting of
the RuO$_6$-octahedra around a non-symmetric axis in the basal plane,
leading to an {\it orthorhombic} crystal symmetry.\cite{friedt}
This results in modified magnetic properties: the Curie-like behavior
is truncated, i.e. the susceptibility decreases after a pronounced
maximum at low temperature. This behavior suggests antiferromagnetic
correlations with a
characteristic temperature scale which gradually increases, from $ T
\sim 0$K at
$ x =0.5 $ to about 10 K as $x$ approaches 0.2.
In addition, a pronounced inplane anisotropy in
the spin susceptibility appears, indicating spin-orbit coupling effects
with a two-fold symmetry axis. For both region II and III the
system remains metallic.  Finally, there is a first-order phase
transition at $ x \approx 0.2 
$ to a phase (region I, i.e. $ 0 \leq x < 0.2 $ ) with flattened
RuO$_6$-octahedra at low  temperatures.  This phase is
continuously connected with the pure 
Ca$_2$RuO$_6$, which is a Mott-insulator and shows
antiferromagnetic order with
$ T_N \sim $ 100 - 150 K. The unusual sequence of changing properties
upon Ca-Sr alloying has motivated a number of theoretical studies.\cite{nomura,terakura,anis1,anis2,hotta}
 
Before going into details of different attempts to explain the
physics underlying these properties we
discuss the basic electronic structure of
the metallic Sr$_2$RuO$_4$, which is well understood experimentally as
well as theo\-re\-tic\-al\-ly.\cite{BERGEMANN,SHEN,LDA} This will be the
basis for all following 
discussions of the alloy. The electronic band structure receives its
basic character from the $4d$-$t_{2g} $ orbitals of the Ru-ion: $ \{
d_{yz}, d_{zx}, d_{xy} \} $ which disperse dominantly in each RuO$_2$-plane
via the $\pi$-hybridization with the O-$2p$-orbitals. It is easy to
see that the $ d_{yz} $- and $ d_{zx} $-orbitals have a very
anisotropic dispersion yielding two quasi-one-dimensional bands, that
hybridize to form one electron-like ($ 
\beta $-band)  and one hole-like ($ \alpha $-band) Fermi surface. The
quasi-one-dimensional nature of these bands produce strong
nesting effects which gives rise to enhanced incommensurate spin
fluctuations around a wave vector $ {\bf Q} \approx (2\pi/3a ,2\pi/3a
, 0) $ \cite{nomura,MAZIN,KKNG}
as observed in neutron scattering ($a$ being the inplane lattice
constant).\cite{SIDIS} The $ \pi $-hybridization with the
intermediate O $2p$-orbitals in both inplane directions generates for the
$ d_{xy} $-orbital a wider
genuinely two-dimensional 
$ \gamma $-band which has van Hove singularities rather close to the
(electron-like) Fermi surface.  In this way the band structure known
from de Haas-van Alphen \cite{BERGEMANN} and ARPES measurements
\cite{SHEN} is qualitatively well understood. 

The doping of Ca for Sr introduces a
gradual rotation of the RuO$_6$-octahedra around the $z$-axis and
yields a modification in the band structure, which the analysis of the
emerging magnetic properties in region III is based on.\cite{friedt}
Both groups, Nomura and Yamada (NY)\cite{nomura}, and Fang and
Terakura (FT)\cite{terakura}, emphasize that
octahedra rotation tends to narrow particularly the $ \gamma $-band
affecting the band structure 
in a way that the $ \gamma $-band van Hove singularity becomes a
dominant feature for the electronic properties. This leads to an
increase of the uniform spin susceptibility in the region
III via the increase in density of states and the Stoner
enhancement. On the other hand, octahedra tilting, as 
occurring in region II, tends to narrow the $ \alpha $-$\beta$-bands
and would shift the nesting vector $ {\bf Q} = (\pi, \pi, 0) $ pushing
the system more towards an antiferromagnetic 
instability.  While this scenario leads to a system with metallic
behavior and an enhanced uniform 
susceptibility in region III, it is difficult to generate the Curie-behavior
displaying apparently one localized $ S=1/2 $ degrees of freedom per
Ru-ion. This spin size is even more surprising in view of the fact
that the fully localized $4t_{2g}$-orbitals with four
electrons on the $ Ru^{4+} $-ion would by Hund's rule favor a spin-1
configuration, as observed in Ca$_2$RuO$_4$. 
Of course, complete localization is ruled out for the system being 
metallic. 

In contrast to the NY and FT models based
on three modified itinerant electronic bands, Anisimov {\it et
al.} focused their 
attention on the emerging spin $ S=1/2 $ in the susceptibility around
$x = 0.5 $.\cite{anis1,anis2} In their
analysis the 
important feature of the RuO$_6$-octahedra rotation is a progressive
narrowing of the bands with increased Ca-doping. Due to the different
character of the three bands (the parity with respect to reflection at
the basal plane is opposite for the $\alpha \beta
$-orbitals and $ \gamma $-orbital), electron
correlation will then drive a Mott transition which is orbital
selective.  Since the $ \alpha $-$\beta$-bands, derived from
quasi-one-dimensional bands, have in Sr$_2$RuO$_4$ approximately half the
width of the genuinely two-dimensional $ \gamma $-band, correlation would
be more effective to localize  their carriers. Using LDA augmented
by dynamical 
mean field theory (DMFT) and  non-crossing approximation (NCA) Anisimov
{\it et al.}  demonstrated that the band narrowing can indeed lead
to a redistribution of the four electrons per Ru-ion and Mott
transition for the $ \alpha $- and $ \beta $-band.\cite{anis1,anis2}
The two orbitals 
$ d_{yz} $ and $ d_{zx} $ absorb 3 electrons while the $ \gamma $-band 
with one electron, although half-filled, remains metallic. The
3 electrons 
on the two orbitals $ d_{yz} $ and $ d_{zx} $ generate a localized 
spin 1/2 and an orbital isospin 1/2 degrees of freedom. This
change of electronic properties is assumed to occur
upon doping and is essentially
reached when $ x= 0.5 $ is approached. For the range $
0.2 < x \leq 0.5 $ the magnetic properties  
are dominated by the combination of orbital and spin degree of
freedom. While this picture results in a localized spin 1/2 per site,
as observed in the spin susceptibility and leads to a slight
elongation of RuO$_6$-octahedra ($ \sim 1 - 2 \% $) as the orbital
occupation is rearranged \cite{friedt}, it seems in contradiction with
other measurements. Recent optical spectroscopy data do not observe as
drastic a change of carrier concentration in the range $ 0.5 < x < 0.9 $
as one would anticipate from an orbital-selective Mott
transition.\cite{LEE} Furthermore, an apparent inconsistency 
in the orbital structure has been reported based on neutron scattering
data.\cite{NEW-BRADEN} Recently, Liebsch argued based on a modified
DMFT approach that orbital-selective Mott transition is unlikely to
occur for the $ 4d - t_{2g} $-orbitals.\cite{liebsch} However, Koga et
al. have very recently demonstrated that under rather general
conditions an orbital-selective Mott transition in a two-band Hubbard
model is possible, contradicting Liebsch's argument \cite{KOGA}.

Despite the mentioned reservation towards the orbital-selective Mott
transition we will use this scenario in this paper 
as a {\it working hypothesis} and investigate some
possible consequences mainly for the region II of the NM phase
diagram. We find properties of the model which are in surprisingly
good qualitative agreement with important features experimentally observed. 
Assuming three localized electrons in the $ 
d_{yz}$-$d_{zx} $-orbitals we discuss the effective 
Kugel-Khomskii-type Hamiltonian
describing the corresponding localized degrees of freedom, spin and
orbital isospin, within a mean field theory. Including doping 
disorder into the model gives an important clue for the comparison
with experimental findings. In particular, it allows to discuss the observed
metamagnetic transition and the unusual magnetoresistance behavior in
region II. We will conclude that the picture of localized $ 
d_{yz}$-$d_{zx} $-orbitals gives a good description of the magnetic
properties of the region II which is probably difficult to explain
with only itinerant electrons in all three orbitals. 

\section{Effective Kugel-Khomskii Model}

For doping concentrations $x$ in region II and
 at the boundary of III the $
d_{yz}$-$d_{zx} $-orbitals are considered to be localized, according to
Anisimov {\it et al.}.\cite{anis1,anis2} We now introduce an effective
model for these localized orbitals, and will neglect the still itinerant $
\gamma $-band (which will play a role again in a later stage).
This is obviously a drastic simplification in view of the influence
 the $ \gamma $-band may have on the two localized 
orbitals. There are two basic
types of coupling: (1) RKKY-interaction and (2) double exchange. 
The former tends towards an antiferromagnetic spin correlation, while
 the latter, in contrast, would prefer ferromagnetism. 
Despite the fact that the $ \gamma $-band is half-filled, the band
structure is far from particle-hole symmetric, so that no
dangerous singular features in the RKKY coupling would appear. 
Thus it is unclear which trend would win. We shelve this problem for
the following discussion, acknowledging the
short-coming of in our simplified model. The considerably more complex
complete model will be a matter of future study.

The three localized electrons in the two orbitals ($ d_{zy}, d_{zx} $)
 create four different 
configurations, consisting of a spin 1/2, $ | \uparrow \rangle $ and
$ |\downarrow \rangle $, and an orbital degree of freedom, $ | + \rangle
$ and $ | - \rangle $ describing the singly occupied $ d_{zx} $ and $
d_{yz} $, respectively.  
Analogous to the spin also the orbital degree of freedom corresponds
to a two-dimensional SU(2)-symmetric Hilbert space represented by an
isospin. We 
define, therefore, isospin operators  $ {I} $ which act 
in the following way
\begin{equation}
I^z | \pm \rangle = \pm \frac{1}{2} | \pm \rangle, \qquad I^+ |-
\rangle = | + \rangle, \qquad I^- | + \rangle = | - \rangle \; .
\end{equation}
Hence, on every site
the localized state consists of a set of four product states 
\begin{equation} 
\{ | \uparrow + \rangle,
\; | \uparrow - \rangle , \; | \downarrow + \rangle , \; |\downarrow -
\rangle \} .
\end{equation}
The fundamental microscopic model providing the dynamics for these
degrees of freedom is an extended two-orbital Hubbard model (neglecting
the $\gamma$-band) where we restrict ourselves to nearest-neighbor
hopping and onsite interaction for the intra- and inter-orbital Coulomb
repulsion, $ U $ and $ U' $, respectively, and the Hund's rule coupling $
J_H$.  

\begin{equation} \begin{array}{ll}
{\cal H}_{\alpha,\beta} = & \displaystyle -t \sum_{i,{\bf a},s}
\{c_{i+a_y,yz, s}^{\dag} c_{i,yz,s} + c_{i+a_x,yx,s}^{\dag} c_{i,zx,s} +
h.c.\}  \\ & \\ & \displaystyle 
- \mu \sum_{i,s,\nu}  c_{i,\nu,s}^{\dag} c_{i,\nu,s} \\ &
\\ & \displaystyle + U
\sum_{i} \sum_{\nu} n_{i\nu \uparrow} n_{i \nu \downarrow} 
+ U' \sum_{i} n_{i,zx} n_{i,yz} \\ & \\ & \displaystyle +J_H
\sum_{i,s,s'} c_{i,yz,s}^{\dag} 
c_{i,zx,s'}^{\dag} c_{i,zx,s} c_{i,yz,s'}  \;,
\end{array}
\end{equation}
where $ c_{m,i,s}^{\dag} $ ($ c_{m,i,s} $) creates
(annihilates) an electron on site $ i $ with orbital index $ \nu $ ($ =
yz,zx $) and spin $ s $ ($ n_{i,\nu,s} = c_{i,\nu,s}^{\dag}
c_{i,\nu,s} $, $ n_{i,\nu} = n_{i, \nu, 
\uparrow} + n_{i, \nu, \downarrow} $; $ {\bf a} = (a_x , a_y ) = (1,0) $
or $ (0,1) $ basis lattice vector). 
The effective nearest-neighbor Hamiltonian in terms of spin and
isospin has the form of a Kugel-Khomskii model and 
can be derived within second order perturbation in $ t/U $,

\begin{eqnarray}
{\cal H} & = & J \sum_{i,{\bf a}} \left[ \left\{A (I^z_{i+{\bf a}} +
    \eta_{{\bf a}})(I^z_{i} + 
\eta_{{\bf a}}) +B \right\} 
{\bf S}_{i+{\bf a}} \cdot {\bf S}_{i} \right.  \nonumber \\
& & \left. + [C (I^z_{i+{\bf a}} +
\eta'_{{\bf a}})(I^z_i + \eta'_{{\bf a}}) + D \right]
\label{Heff}
\end{eqnarray}
with the coefficients 
\begin{eqnarray}
A &=& \frac{3 \alpha^2 + 1}{(3 \alpha-1)(\alpha +1)}  \label{pm1} \\
B &=& \frac{-(1- \alpha)^2}{(3 \alpha^2 + 1)(3 \alpha-1)(\alpha +1)}
\label{pm2} \\
C &=& \frac{5 - 3\alpha}{4 (3 \alpha - 1)} \label{pm3}  \label{CC} \\
D &=& \frac{1}{(5 - 3 \alpha)(3 \alpha - 1)} \label{pm4}  \\
\eta_{{\bf a}} &=&  \frac{(3 \alpha - 1)(\alpha + 1)}{2(3 \alpha^2
  +1)} (a_x^2 - a_y^2) \label{pm5} \\
\eta'_{{\bf a}} &=& \frac{3\alpha-1}{2(5 - 3\alpha)} (a_x^2 - a_y^2)
 \label{pm6} 
\end{eqnarray}
where $ \eta_{{\bf a}} $ and $ \eta'_{{\bf a}} $
have opposite sign for the $x$- and $y$-axis bonds. 
We impose the (approximatively valid) 
relation $ U = U' + 2 J_H $ with $ \alpha = U'/U $ and
$ J = 4t^2/U $. In order to obtain a
valid approximation $ \alpha $ should lie between 1/3
and 1, and not too close to these 
boundary values. We choose throughout this
paper $ \alpha = 3/4 $ as a representative value whenever we do a
concrete analysis of this model.  We emphasize that while
the spin has complete SU(2) symmetry, the isospin has only Ising-like
interactions, i.e. quantum fluctuations are suppressed for the latter. 

The Hamiltonian (\ref{Heff}) describes the system with tetragonal
crystal symmetry corresponding to regime III in
the experimental phase diagram close to $ x= 0.5 $. Now we add terms which
describe the orthorhombic distortion due to the tilting of the
octahedra with the rotation axis in the basal plane. We do not
speculate about the origin of this lattice deformation, which we consider to
lie outside our model. Following
Ref. \cite{anis2} we introduce these with the strains
$ \epsilon_1 = \epsilon_{xx} - \epsilon_{yy} $ corresponding to a
rotation axis like [100] and $ \epsilon_2 = \epsilon_{xy} $
corresponding to the axis [110]. These distortions lift the
orbital degeneracy of $ (d_{yz}, d_{zx} ) $ and, hence, can be viewed
as an
effective ``uniform field'' polarizing the isospins in a certain
direction. From symmetry considerations it follows that

\begin{equation}
{\cal H}_{dist} =  \sum_i [K_1 \epsilon_1 I^z_i + K_2 \epsilon_2 I^x_i]
\label{CF-1}
\end{equation}
where $ K_{1,2} $ are coupling constants (see the Appendix). The first term is
a longitudinal field while the 
second term corresponds to a transverse field, introducing quantum
fluctuations. In connection with the 
Ising Hamiltonian (\ref{Heff}) this second term provides the possibility of a
continuous quantum phase transition.\cite{sachdev} The lattice
distortion in the regime II has a tilt axis lying between [100] and
[110],  so that both strains are turned on, if the border at $ x
=0.5 $ between the two phases is crossed. 

\section{Mean field approximation}

We now discuss the basic properties of this model and its phase
diagram within a mean field approximation
for both the spin and isospin degree of freedom. The Hamiltonian 
on the square lattice possesses a bipartite structure, which is the
basis of our mean field decoupling. We introduce different mean fields
for the corresponding $A$- and $B$-sublattice.

\begin{equation}
\langle I^z_i \rangle = \left\{ \begin{array}{ll} t_A & i \in A \\
t_B & i \in B \end{array} \right. \quad \mbox{and} \quad 
\langle S_i^z \rangle = \left\{ \begin{array}{ll} s_A  & i \in A \\
s_B & i \in B \end{array} \right.
\end{equation}
The mean field calculation is straightforward, whereby the self-consistent
equations are solved numerically. We find the following
properties.

\subsection{Tetragonal system}

First we consider the situation in the absence of any orthorhombic 
strains, $
\epsilon_1 = \epsilon_2 = 0 $. For the choice of parameters ($ \alpha
= 3/4 $) we find
that the coupling of the isospins, represented by $JC$, has a higher
energy scale than the one for the spins, leading to a higher mean field
transition temperature. The parameter $ C $ in
Eq.(\ref{Heff},\ref{CC}) is positive, favoring a
staggered (Ising-antiferro-) orbital order. 
This orbital order implies ferromagnetic
exchange between the spins. Note that this gives rise to enhanced
uniform spin 
susceptibility as observed in experiment.
The transition temperature to ferromagnetic order is
low compared to the energy scale of the orbital order, 
so that it is natural to expect over a wide range of temperature a
Curie-like susceptibility originating from almost independent spin 1/2
degrees of freedom, consistent with the experimental observation. 

\subsection{Orthorhombic distortion}

Now the uniaxial strains $ \epsilon_1 $ and $ \epsilon_2
$ are turned on. Since in regime II  
the crystal symmetry is determined by the rotation of the octahedra around
an inplane axis which does not coincide with a crystal symmetry axis,
both strains, $ \epsilon_1 $ and $ \epsilon_2 $ appear. For our
calculations we assume that 
$ K \epsilon_1 = 0.75 K
\epsilon_2 = K \epsilon $ which we use as a control parameter for our phase
diagram. Let us first consider the orbital part only, ignoring the
spin part. We observe a competition between the
staggered order and the uniform alignment induced by the strain
field. The transverse isospin component $ I^x $ 
induced by the strain $ \epsilon_2 $ reduces the transition
temperature, driving the system towards a quantum
phase transition. This is included in our mean field approach with the
following self-consistence equations.

\begin{equation} \begin{array}{ll}
t_A = & \displaystyle - \frac{2 JCt_B + K_1 \epsilon_1}{2 \sqrt{(2JCt_B + K_1
    \epsilon_1)^2 + K_2^2 \epsilon_2^2)}} \\ & \\
& \displaystyle \times  \tanh \left( \beta \sqrt{(2JCt_B + K_1
    \epsilon_1)^2 + K_2^2 \epsilon_2^2)} \right).
\end{array}
\end{equation}
The isospin component $ t_x = \langle I^x \rangle $ is obtained
as

\begin{equation} 
\begin{array}{ll} 
t_{xA} = & \displaystyle  - \frac{ K_2 \epsilon_2}{2 \sqrt{(2JCt_B + K_1
    \epsilon_1)^2 + K_2^2 \epsilon_2^2)}} \\ & \\
& \displaystyle \times  \tanh \left( \beta \sqrt{(2JCt_B + K_1
    \epsilon_1)^2 + K_2^2 \epsilon_2^2)} \right)
\end{array}
\end{equation}
for the $A$-sublattice site. The analogous expressions for the $
B$-sublattice site are obtained by exchanging $ t_B $ and $ t_A $ in
these equations. 
The $ t_x$-component is only finite, if the orthorhombic strain $
\epsilon_2
$ is present. 
\begin{figure}
\resizebox{0.4\textwidth}{!}{%
  \includegraphics{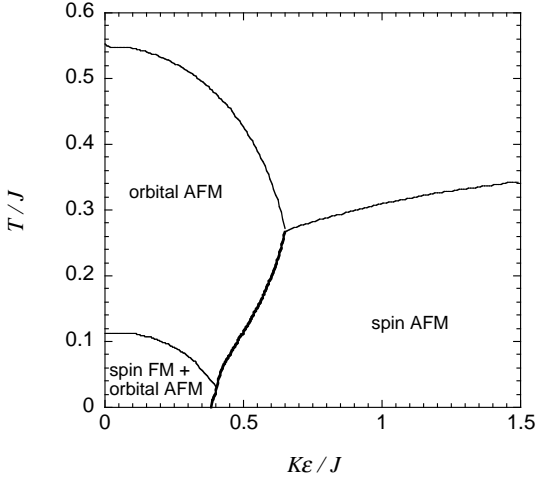}}
\caption{Mean field phase diagram as a function of temperature $T$ and
  strain $K \epsilon $. Thin lines denote second order phase
  transitions and the thick line a first order phase transition.} 
\label{fig:phase}
\end{figure}
The staggered orbital moment $ t_s= (t_A -
t_B)/2 $. Increasing the strains leads to a gradual diminishing of $
t_s $ and the corresponding transition temperature until it would vanish
at a quantum 
critical point. In turn the mean fields $ (t_{xA}+t_{xB})/2 $ and
$ t_0= (t_A + t_B)/2 $ increase continuously favoring AF spin
correlation resulting in the phase diagram shown in
Fig. \ref{fig:phase}. The decrease of $ t_s$ leads at the same time to a
weakening of the ferromagnetic correlation and the gradual increase of a
uniform isospin correlation gives rise to a competing
antiferromagnetic spin correlation. 
Indeed the continuous evolution of $ t_s $ is truncated
by the onset of antiferromagnetic spin order,
before the anticipated quantum suppression of the staggered isospin
order is completed. This
spin correlation is favored by the strain driven ferro-orbital correlation
and has a higher energy scale (mean field transition temperature)
than the ferromagnetic order. Consequently, the staggered isospin phase
and the  antiferromagnetic order are separated by a first order
transition. There is even a
reentrant behavior associated with this first order instability such that
around a critical range of $ K \epsilon \approx 0.5 J $  upon lowering
temperature first  a staggered isospin phase with enhanced tendency
towards ferromagnetism would be reached and with a first order
transition (thick line in Fig. \ref{fig:phase}) 
at lower temperature  the antiferromagnetic phase would appear. 
Otherwise all phase transitions in the phase diagram of Fig.
\ref{fig:phase} are second order (thin lines). 

The evolution of the system, under orthorhombic distortion, from 
ferromagnetic towards antiferromagnetic behavior displays the correct
qualitative trend in comparison with the experimental situation. 
The fact that the temperature scale of the antiferromagnetic state is
higher than that of the ferromagnetic is also in qualitative
agreement with experimental data. The former is easily visible
in experiment at a 
temperature $ T \sim 5 - 10 K $ around $ x = 0.2 $, while the latter can
brought into connection with a irreversibility transition observed in
the vicinity of $ x =0.5 $ below $ \sim 1 K $.\cite{SUGAHARA} The
irreversibility and 
the sensitivity to slow dynamics led to the interpretation as a cluster
glass. Similar features could arise in an inhomogeneous
ferromagnetic phase.\cite{SUGAHARA}

\subsection{Spin-orbit coupling and anisotropic spin susceptibility}

For the Ru-ion spin-orbit coupling is not negligible. The microscopic
formulation of spin-orbit coupling involves the whole set of
$t_{2g}$-orbitals including the itinerant $ \gamma $-orbital.\cite{kong}
We remain within our present reduced model and restrict
ourselves to a phenomenological approach where the effect of
spin-orbit coupling enters via the $ g $-tensor modifying the
Zeeman term in the Hamiltonian. This term can be derived from symmetry
considerations analogous to the coupling of the strain (Appendix A). 
\begin{equation} \begin{array}{l} \displaystyle 
{\cal H}_Z = - \mu_B \sum_{i} [ g {\bf H} \cdot {\bf S}_i +  g_1
   \langle I^z_i \rangle  
  (H_x S^x_i - H_y S^y_i) \\ \\ \displaystyle \qquad \qquad + g_2
   \langle I^x_i \rangle ( H_x S^y_i + H_y S^x_i )]
\end{array}
\end{equation}
where only the two isospin directions $ I^z $ and $
I^x $ have been included, ignoring $ I^y $ which is not induced by strain nor
interaction ($ \mu_B $: Bohr magneton). Two coupling
constants $ g_1/g $ and $g_2/g $ enter  
as phenomenological parameters. $ \langle I^a \rangle $ denotes the
average of the orbital configurations on all sites. 
Here we want to concentrate on the
anisotropy of the susceptibility in the basal plane ($x$-$y$) which
shows the most distinguished effect. 

We now calculate the spin-susceptibility on the background of the 
staggered orbital order and the driven ferro-orbital correlation:
\begin{equation}
\chi(T,\phi) = \chi_0(T) \{1 + \frac{g_1}{g} \langle I^z \rangle \cos 2
\phi + \frac{g_2}{g} \langle I^x \rangle \sin 2 \phi \}
\end{equation}
where the angle corresponds to the orientation of the field in the
basal plane relative to the $x $-axis. The isotropic susceptibility $
\chi_0 (T) $ includes the spin correlation and approaches the behavior
of free spins for temperatures much higher than the Ne\'el temperature
in our model. 

\begin{figure}
\resizebox{0.4\textwidth}{!}{%
  \includegraphics{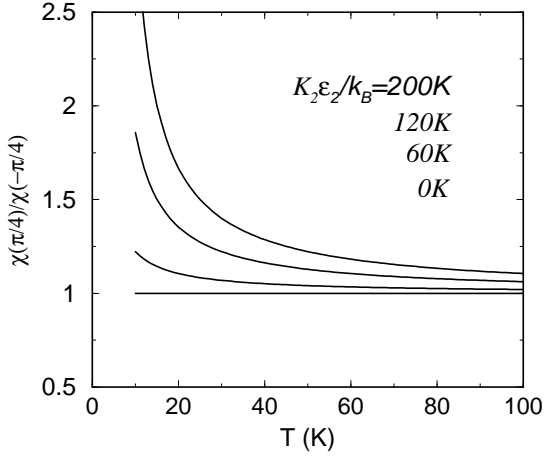}}
\caption{Anisotropy of susceptibility: Ratio of the maximal and
  minimal susceptibility as function of temperature for $ g_2/g = 0.1
  $ and different levels of distortion $ K_2 \epsilon_2 $.} 
\label{fig:anis}
\end{figure}

In the tetragonal phase where a staggered Ising orbital order is
realized the average of the isospin vanishes and, consequently, there
is no basal plane anisotropy in the spin susceptibility.
This is fully consistent with the experiment at $ x =0.5 $. 
On the other hand, the orthorhombic distortion which yields polarized
orbitals in region II gives rise to anisotropy. Let us for simplicity
now assume that the orthorhombic distortion is of the type $
\epsilon_{xy} \neq 0 $ and $ \epsilon_{xx} - \epsilon_{yy}=0 $. 
Hence we find that only $ \langle I^x \rangle \neq 0 $. If the
temperature is much higher than the characteristic energy scale for
the orbital correlation, the isospin can be approximately considered
as an independent degree of freedom in a driving field, i.e. $ \langle I^x
\rangle \approx K_2 \epsilon_{xy} / 4 k_B T $. Here we assume that the
strain $\epsilon_{xy} $ only weakly depends on temperature and keep it
constant. This is only valid far away from the phase boundary of the
structure phase transition. Thus, the anisotropy of
the spin susceptibility can be described by the ratio of the maximal to
minimal value
\begin{equation} 
\frac{\chi(\pi/4)}{\chi(-\pi/4)}  = \frac{g + g_2 \langle I^x
  \rangle}{g - g_2 \langle I^x \rangle} 
\approx \frac{4 g k_B T + K_2 \epsilon_{xy} g_2}{4 g k_B T - K_2
  \epsilon_{xy} g_2} 
\end{equation}
which tends to 1 for large temperatures and grows for decreasing 
temperatures. Obviously the anisotropy increases too, if $
\epsilon_{xy} $ becomes larger, which implies a growing anisotropy for
samples with $ x \to 0,2 $ in region II. 
For a qualitative view of the behavior we plot for different
parameters $ K_2 \epsilon_2 $ (Fig.\ref{fig:anis}). This behavior is
close to the observed one by NM.\cite{NAKAMAE}

\section{Effects of an external magnetic field and alloy inhomogeneity}

\begin{figure}
\resizebox{0.4\textwidth}{!}{%
  \includegraphics{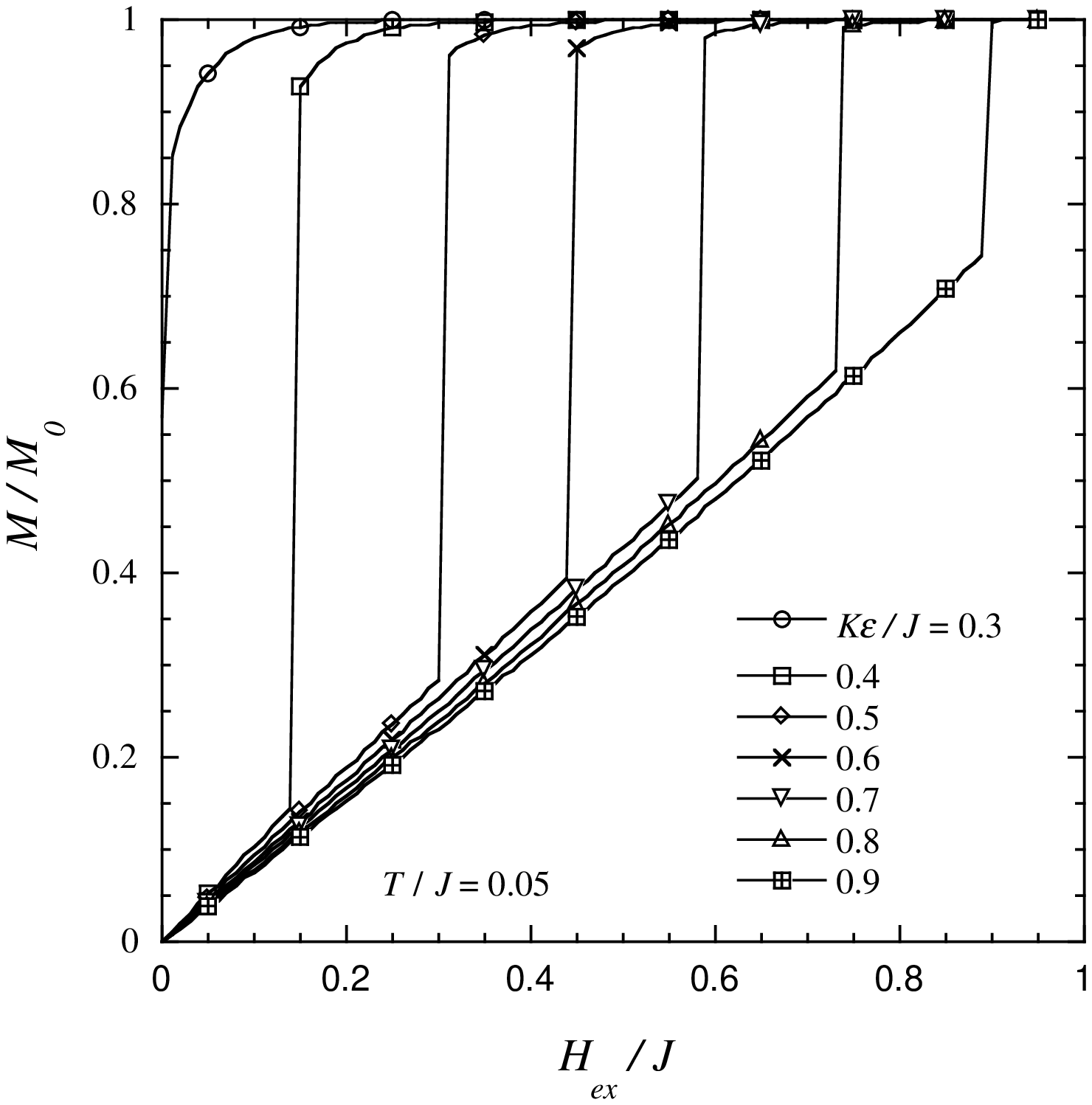}}
\resizebox{0.4\textwidth}{!}{%
  \includegraphics{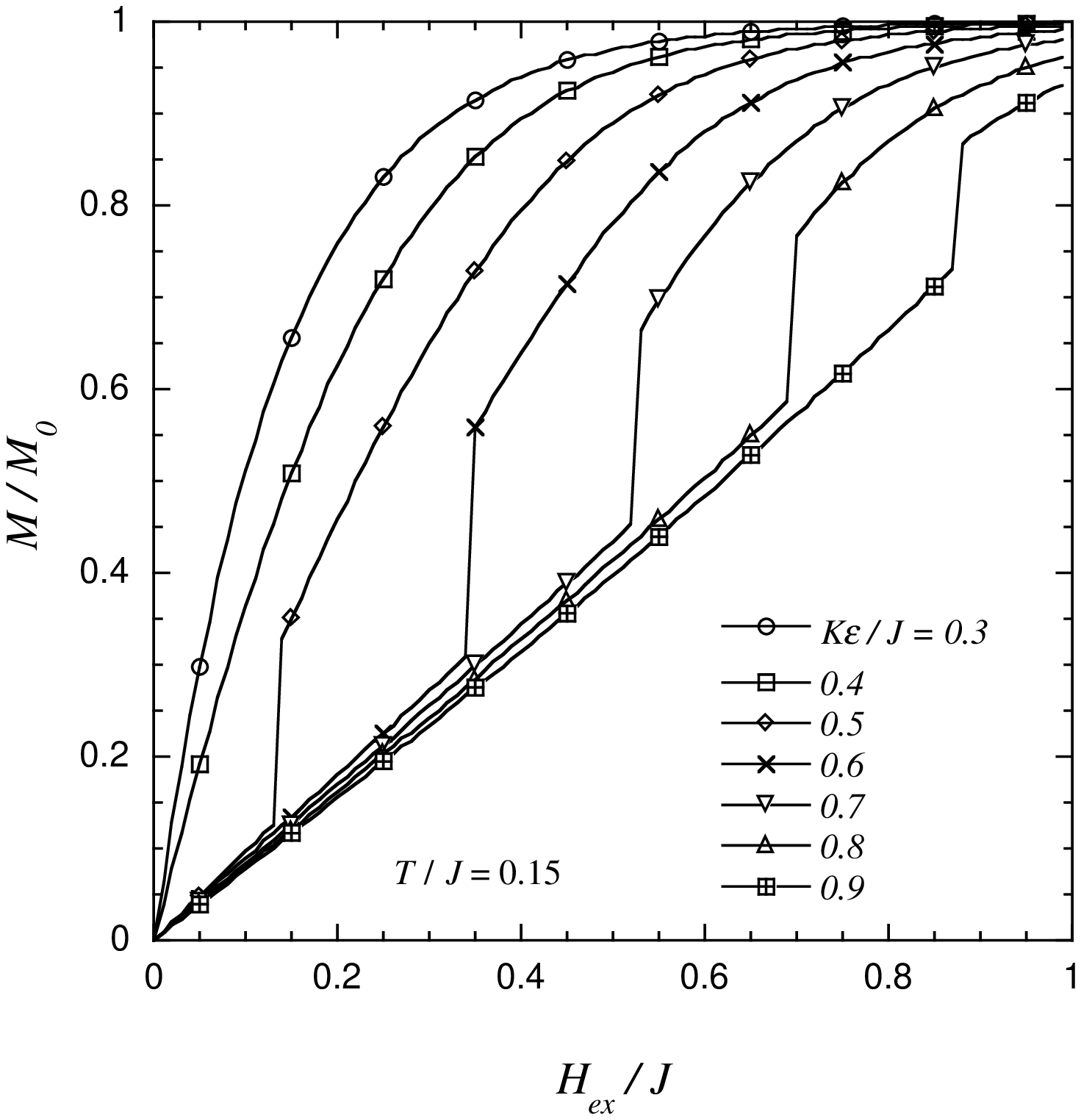}}
\resizebox{0.4\textwidth}{!}{%
  \includegraphics{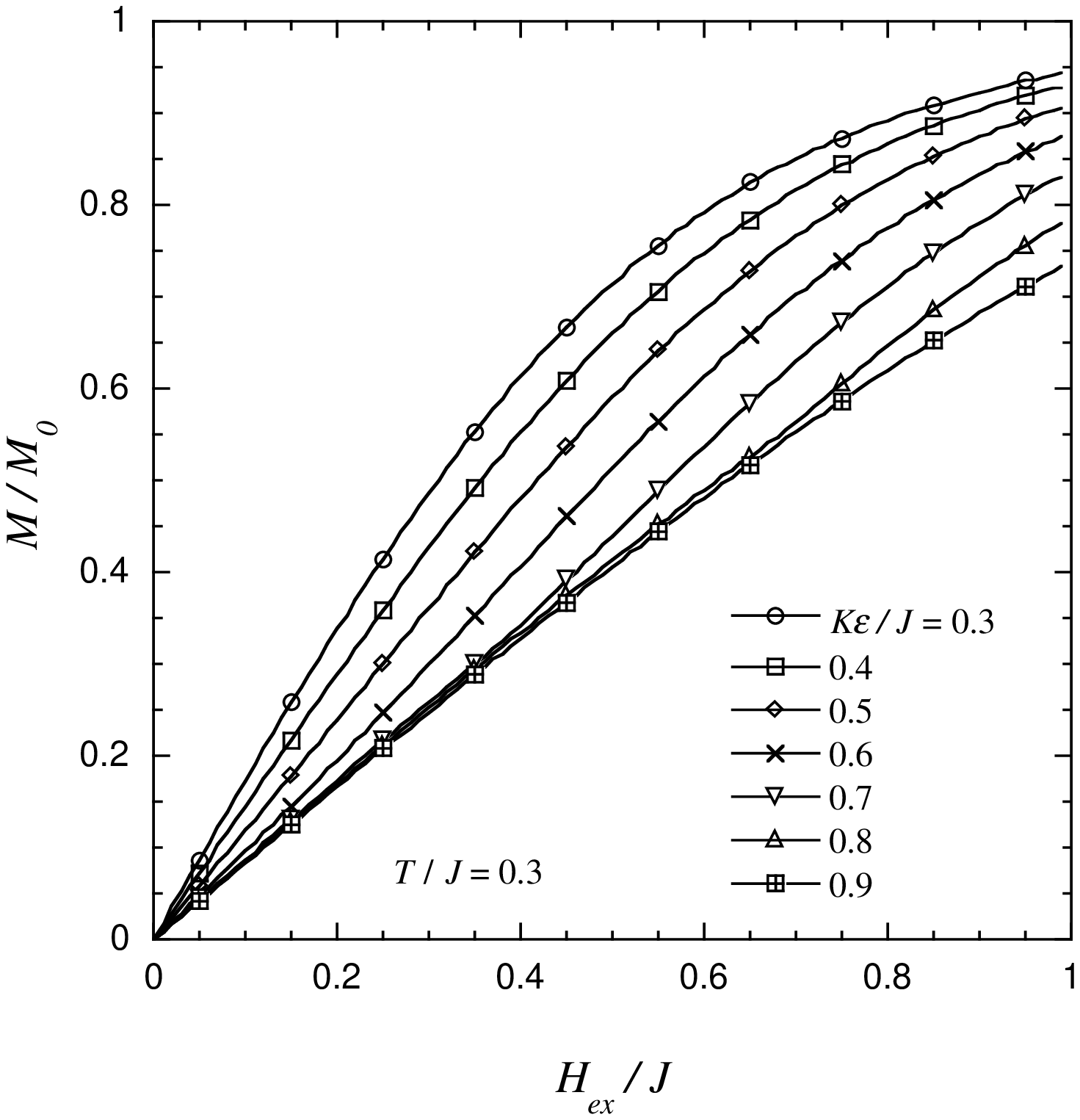}}
\caption{Magnetization curves at three temperatures for different
  values of $K\epsilon/J$.} 
\label{fig:mag}
\end{figure}

We focus now on the phase in which the lattice distortion induces 
the FO-order and AF-spin order. Close to the first-order phase
transition line the phase with a staggered orbital order and the
ferromagnetic spin correlation forms a metastable states. This
can be stabilized by an external magnetic field, since the
antiferromagnetic phase has a lower uniform spin susceptibility. A
discontinuous  ``metamagnetic'' transition appears, since the
magnetic field allows us to traverse the first-order phase transition
line and to reach the phase of higher magnetization. This type of
transition had been investigated in another context by Khomskii and
Kugel \cite{KK}. 
In our mean field
calculation we apply the magnetic field along the $z$-axis and introduce
two additional mean fields
$s_{xA}$ and 
$s_{xB}$ as order parameters of the antiferromagnetic order perpendicular to
the field, in addition to the uniform component along the field.
In Fig. \ref{fig:mag} we show the mean field results of the
magnetization curves 
for three temperatures and a series of distortion strengths. For low 
temperature indeed a discontinuous 
transition between a low and high magnetization
state is found. Obviously, this transition would be hysteretic since
it is first order. Obviously, the field-induced first-order transition
leads to a rearrangement of the orbitals and should result in a 
structural deformation as well \cite{KK}. 

\subsection{Metamagnetic transition}

The real material does not exhibit a discontinuous
metamagnetic transition even at low temperatures. The short-coming in
our model lies in the 
assumption of homogeneity. Obviously, 
the random alloy would rather have spatial 
fluctuations, for example, in the crystal deformation field $ K
\epsilon $. A spatial variation of $ K \epsilon $ would give rise to the
appearance of both the antiferromagnetic and the
staggered orbital phase with ferromagnetic spin fluctuations  
in form of domains. They are separated by rather sharp domain 
boundaries, if the randomness is sufficiently broad and smooth.\cite{YMRI}
We  
distinguish $ A $- and $ F $-domains which denote the domains with
antiferromagnetic and ferromagnetic spin correlation, respectively. 
We consider a system where the antiferromagnetic phase
dominates and the F-domains form at most 
small islands, as would be likely the case for $ x \approx 0.2 $ based
on our model. These $F$-islands develop a magnetic
moment at low temperature once the ferromagnetic correlation length
reaches their spatial extension. In turn the moments of different
islands are correlated via the antiferromagnetically correlated
background. This effective interaction may be frustrating so that
under certain conditions at low enough temperature even glass-like
behavior can occur. We will here, however, concentrate on the
behavior under a magnetic field.

The external magnetic field $ H_{ex} $ polarizes the magnetic moments 
of both domains. Because the F-domains possess the larger
susceptibility, their free energy density drops more rapidly with
increasing field than that of the
A-domains. Thus the F-domains grow at the
expense of the A-domains. The initial linear response of the
magnetization to the field turns nonlinear, if the domains change their
size.
In this way the metamagnetic transition, which was discontinuous and
hysteretic in the homogeneous system, would be smeared, continuous and
reversible for the inhomogeneous case.\cite{YMRI} The inflection point of the
magnetization as a function of $ H_{ex}
$,  would roughly correspond to the highest density of 
the domain boundaries through the system, because this yields the
most rapid change of the domain sizes as a function of $ H_{ex} $.
The metamagnetic transition lies, consequently, also rather close to
the point where the domain boundaries percolate throughout the
sample.

\subsection{Extended mean field approach}

\begin{figure}
\resizebox{0.4\textwidth}{!}{%
  \includegraphics{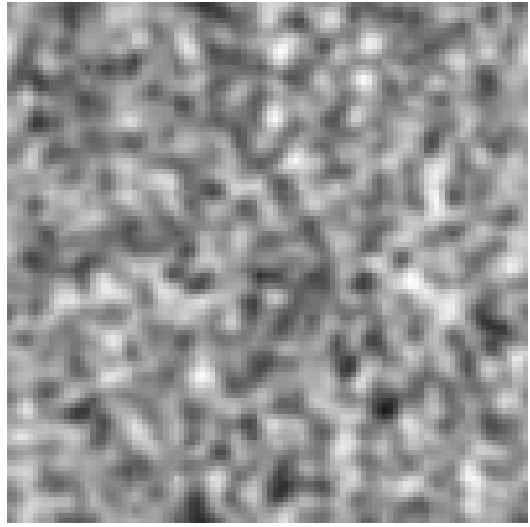}}
\caption{Distribution of $K \epsilon /C$ used in the mean field
  calculation. The 
  mean value of $K \epsilon/C$ is 
$0.9$ and the standard deviation $0.2$. Black corresponds to the minimum value 
of 0.2 and white to the maximum value of 1.6}
\label{fig:dist}
\end{figure}

The continuous metamagnetic behavior of 
the inhomogeneous system can be very easily simulated within our
mean field calculation. We modify Eq.(\ref{CF-1}) to a random field 
coupling

\begin{equation}
{\cal H}_{dist} = \sum_i [K_1 \epsilon_{1i} I^z_i + K_2 \epsilon_{2i}
I^x_i]
\label{CF-2}
\end{equation}
so that for every site, in principle, the crystal field may be different. 
 Since one has to consider certain realizations of
randomness we can only deal with finite lattices. We use a lattice size of
 $100\times100$ to simulate the evolution of the domain distribution in
 a magnetic field. For this
purpose the model was set up in the following way. In order to
obtain a smoothly varying distribution we start with an uncorrelated
uniform distribution of random numbers in the interval $[-0.9,2.7]$
Averaging the values five times with the values at the four nearest
neighbor sites we finally arrive at a smooth distribution shown in
 Fig. \ref{fig:dist}, 
with  mean value $\overline{K \epsilon}/J=0.9$ and with a 
standard deviation $\Delta K\epsilon/j=0.2$, correlated over several
 lattice sites.

For the mean field calculation we start from an initial configuration that
has staggered orbital order for the $t_z$ mean fields, uniform magnetic 
order for the $s_z$ mean fields in the direction of the magnetic field and 
staggered antiferromagnetic order for the $s_x$ mean fields perpendicular
to the magnetic field. In order to find a zero-temperature 
mean field solution we then perform  $10^6$ updates to this mean
field, solving the  
mean field equations each time for one randomly chosen spin, using a quantum operator library developed by one of the authors \cite{QET}.

\begin{figure}
\resizebox{0.4\textwidth}{!}{%
  \includegraphics{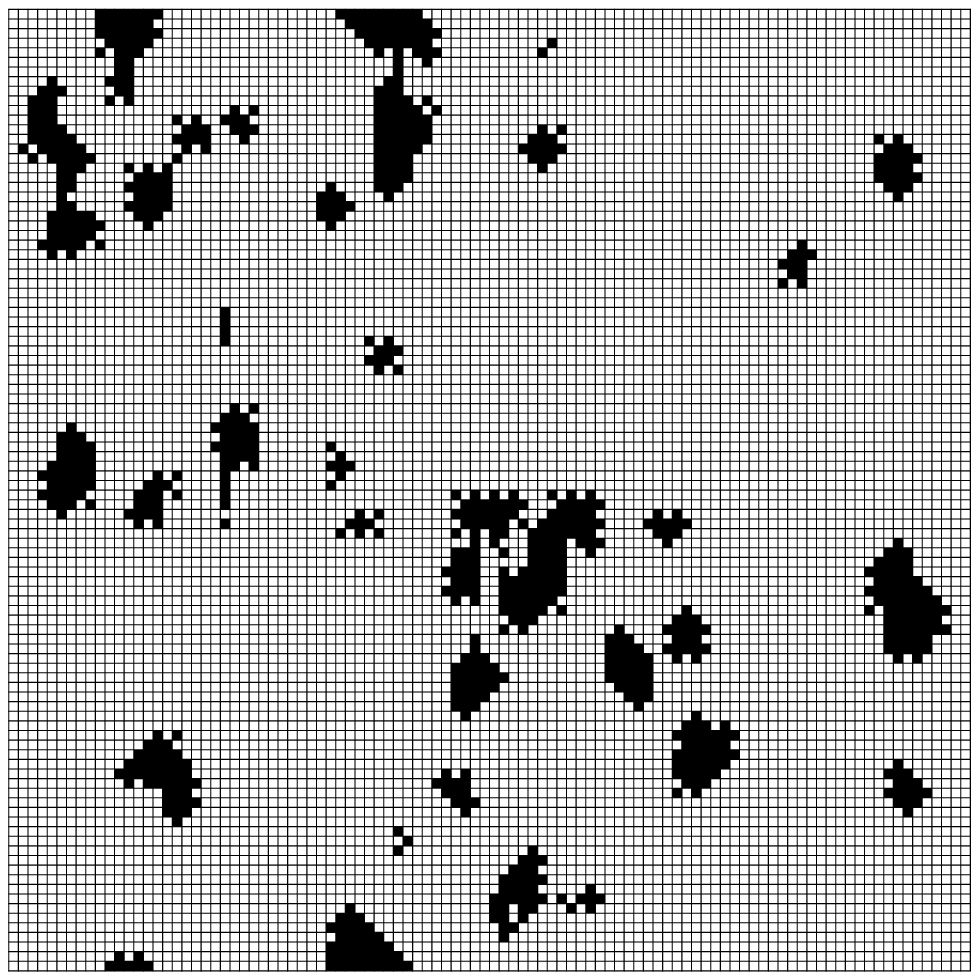}}
\resizebox{0.4\textwidth}{!}{%
  \includegraphics{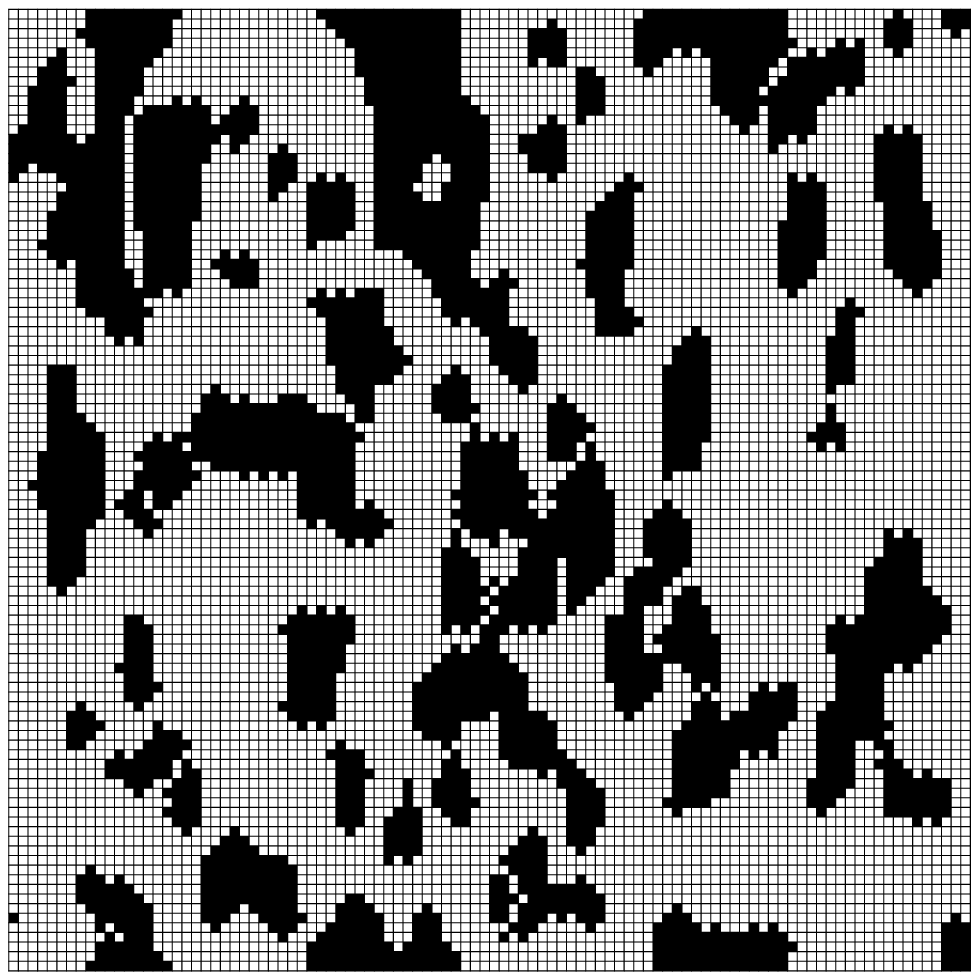}}
\resizebox{0.4\textwidth}{!}{%
  \includegraphics{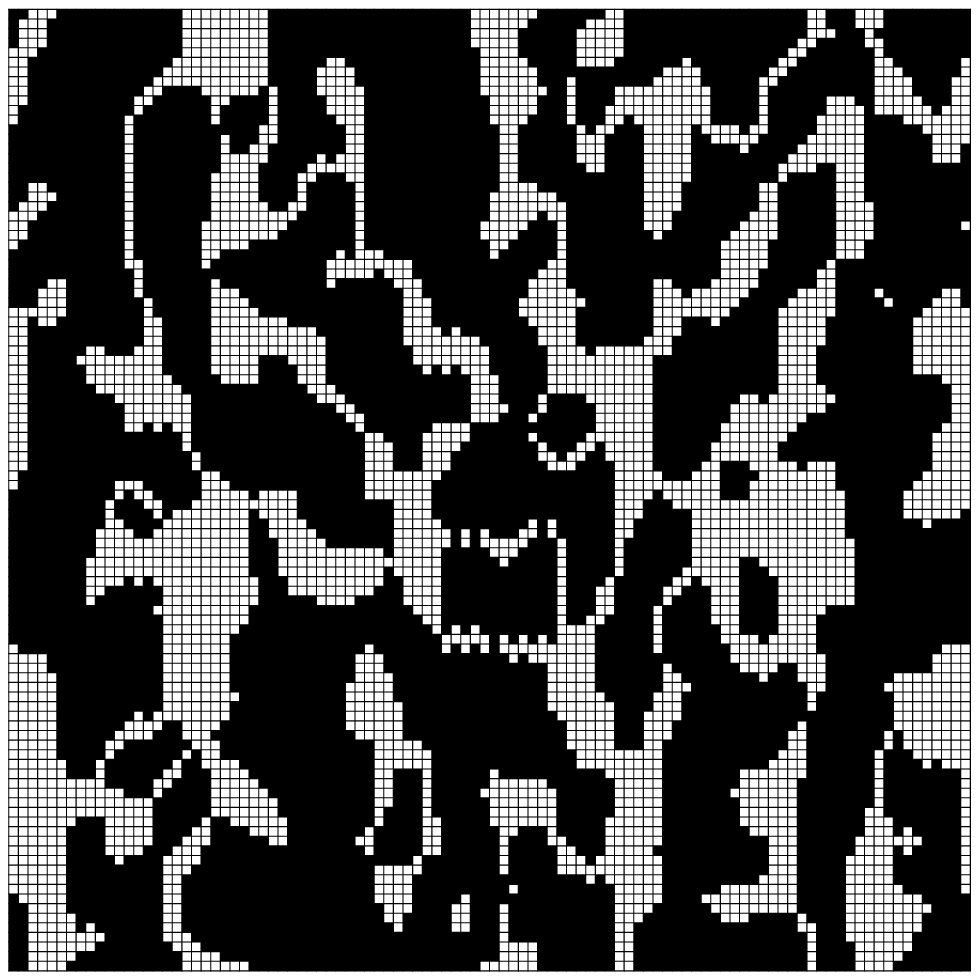}}
\caption{Distribution of ferromagnetic (black) and antiferromagnetic
  (white) domains at a) a weak field $h/J=0.2$ 
and b) close to the percolation threshold with $h/J=0.3$ and c) at
stronger fields $h/J=0.4$.} 
\label{fig:order}
\end{figure}

\begin{figure}
\resizebox{0.4\textwidth}{!}{%
  \includegraphics{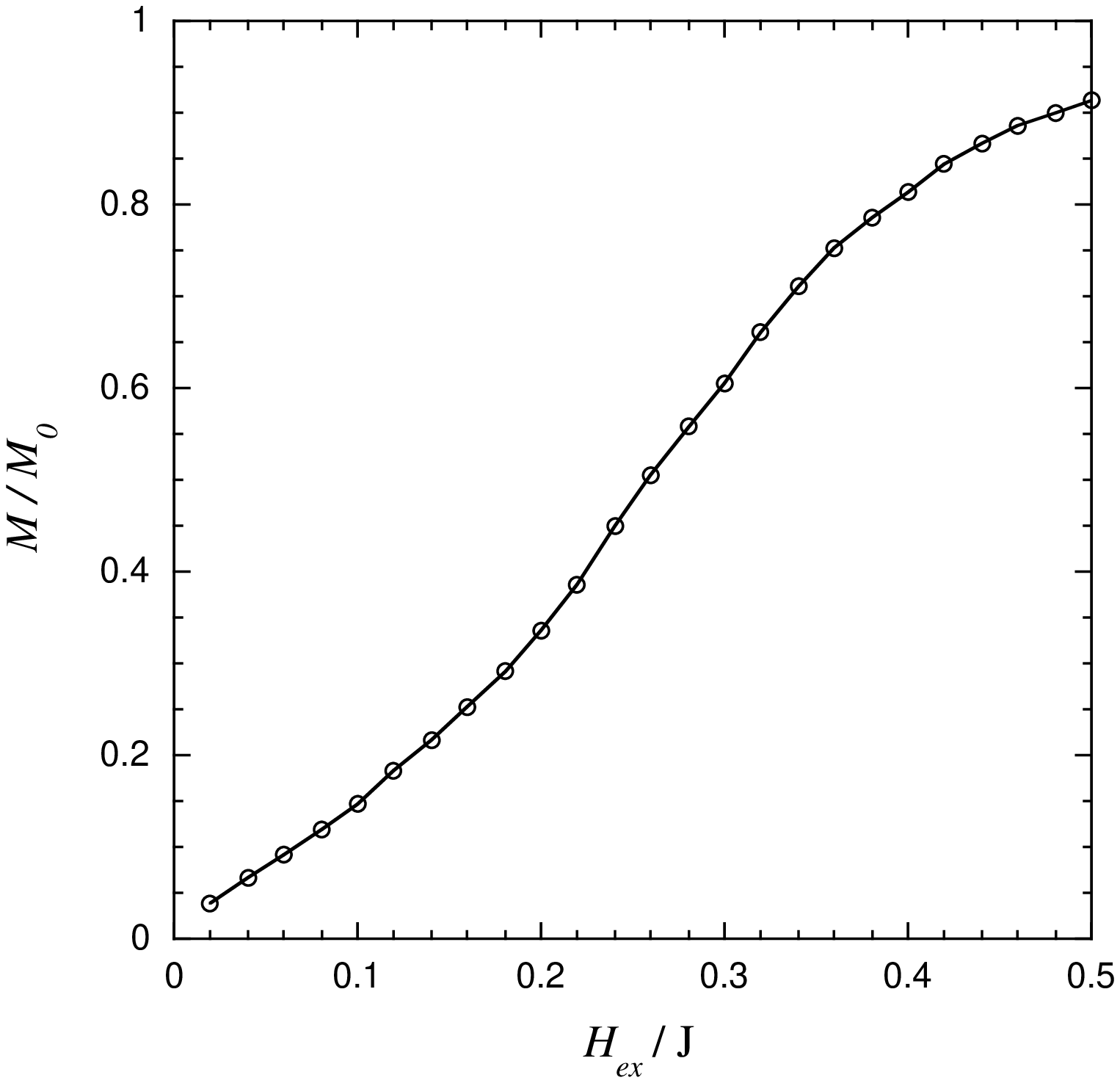}}
\resizebox{0.42\textwidth}{!}{%
  \includegraphics{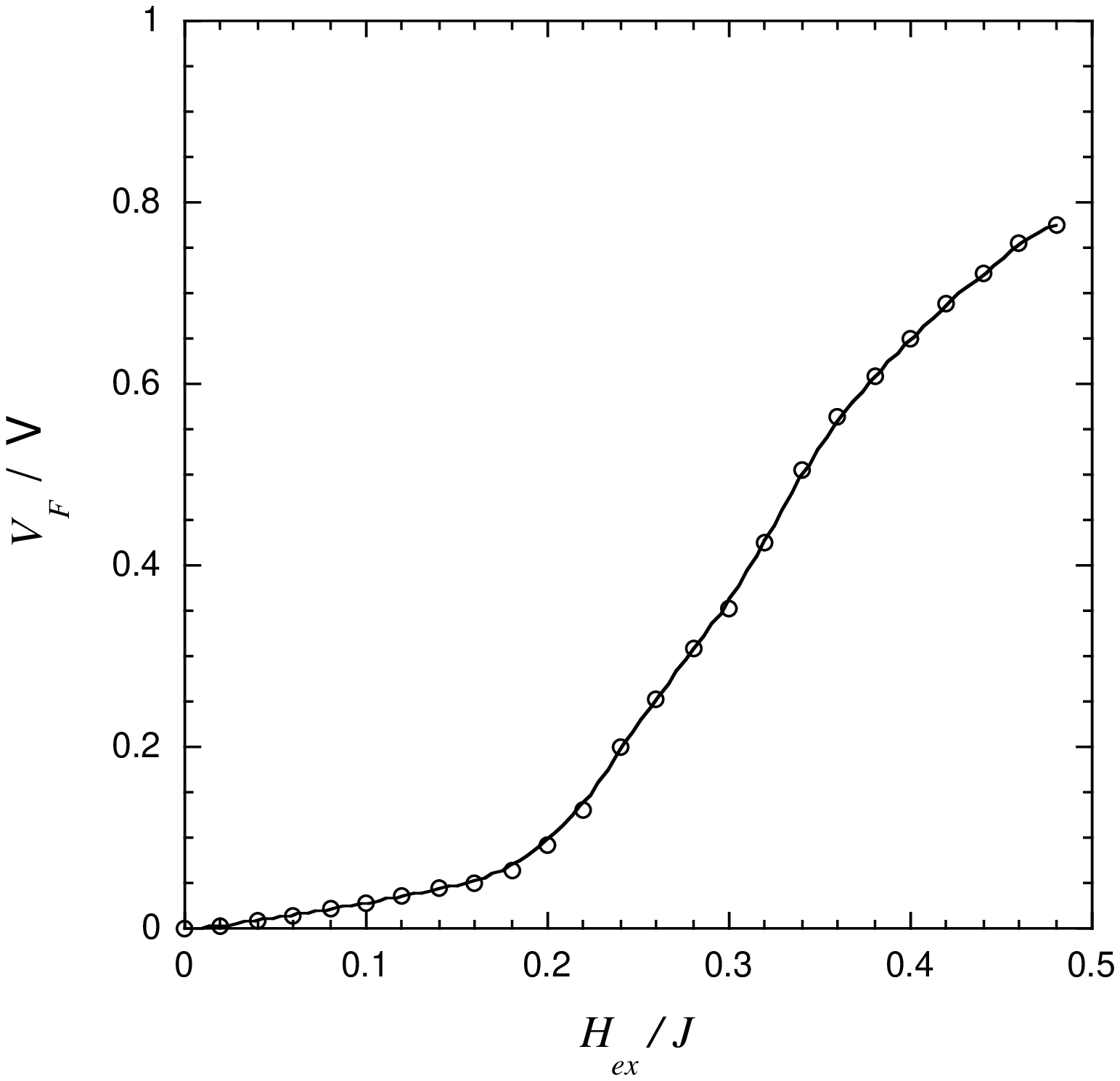}}
\caption{Zero temperature magnetization curve and volume fraction of
  the ferromagnetic domain of the inhomogeneous system with the 
distribution of Fig. \ref{fig:dist}.}
\label{fig:magrand}
\end{figure}

Since most of the $K \epsilon$ values are inside the antiferromagnetic phase
we find only small islands of staggered orbital order and ferromagnetism 
(Fig. \ref{fig:order}a). Increasing the fields we can observe that
these ferromagnetic  
domains grow, and percolate at a ``transition'' to a
ferromagnetic  
phase (Fig. \ref{fig:order}b,c). This transition is accompanied by a smooth
metamagnetic like 
magnetization curve, shown in Fig. \ref{fig:magrand}a. The volume of
the F-domains grow monotonically as shown in Fig.\ref{fig:magrand}b.  

We would like to mention here that the presence of a metamagnetic
transition has been observed in region II \cite{naka-phd,SUGAHARA}. 
Also a structural response has probably be seen in experiment
\cite{FRIEDT2}.

\subsection{Phenomenological model} 

We now illustrate the result of our mean field
simulation by means of a simple phenomenological model describing the
distribution of the F- and A-domains. We define the free
energy at a given temperature as

\begin{eqnarray}
F (l; H_{ex},T) &=& \frac{W}{2} (l - l_0)^2 - V_F(l) f_F(H_{ex},T) -
\nonumber \\ && V_A(l) 
f_A(H_{ex},T),
\end{eqnarray}
where we denote the integrated area of the F- and A-domains by $
V_F(l) $ and $ V_A(l) $ (note $ V_F + V_A = V $ the total area), and
$ l $ is a variable determining the domain distribution. It is
chosen so that the energy expense by deviating from the zero-field equilibrium
state (characterized by $ l= l_0 $) 
follows the quadratic behavior with an ``elastic constant'' 
$ W $. Increasing $ l > l_0 $ leads to the enlarging of $ V_F (l)
$ which is a monotonically growing function of $ l $. The free energy
densities in the two domains are denoted as $ f_F $ and $ f_A $ and
are a function of the field. 

The equilibrium value of $ l $ for a given external field $ H_{ex} $ 
is obtained by
minimizing the free energy. This leads to the 
equation 

\begin{equation} 
W(l-l_0) + [f_F (H_{ex},T) - f_A (H_{ex},T)] \nu(l) = 0
\end{equation}
\label{equilibrium}
with

\begin{equation}
\nu(l) = \frac{\partial V_F}{dl} (l),
\end{equation}
a measure for the density of domain boundaries. 
It is obvious that $ \nu(l) $ would be non-monotonic with a maximum
at an intermediate value of $ l $.  The integral $ \int dl 
\nu(l) $ running from $ l = - \infty $ to $ + \infty $ gives the total
volume $ V $.  This function characterizes the inhomogeneous
system within this theory. Since $ f_F < f_A $ in a finite field
the solution of Eq.(\ref{equilibrium}) 
yields $ l(H_{ex} ) $ as a monotonically growing function of $ H_{ex} $.
The magnetization of the system is given by

\begin{equation} \begin{array}{ll} \displaystyle 
M = - \frac{d F}{dH_{ex}} & \displaystyle = - V_F(l) \frac{d f_F}{d H_{ex}} + 
(V -  V_F(l)) \frac{df_A}{d H_{ex}} \\ &  \\
& = V_F(l) M_F + (V-V_F(l)) M_A \;.
\end{array}
\end{equation}
We use the following simple approximative forms for the free energy
densities in a magnetic field:

\begin{equation}
f_a (H_{ex},T) = f_{0a} - \frac{M_{0a}^2}{\chi_{0a}} {\rm ln} \cosh (
\chi_{0a} H_{ex} / M_{0a} )  
\end{equation}
with $ a = F,A $ and $ f_{0a} = f_a(0) $ ($ f_{0F} = f_{0A} $). Here $
M_{0a} $ is the  
saturation moment and $ \chi_{a0} $ is the uniform spin susceptibility in the
linear response regime at the given temperature $T$. 
Thus, the magnetization is given by

\begin{equation}
M_{a} = - \frac{df_a}{dH_{ex}} = M_{0a} \tanh ( \chi_{0a} H_{ex} / M_{0a} )
\end{equation}
Note that in this discussion we do not rely on the presence of long
range order, neither FM nor AFM. 
The characteristic function $ \nu (l) $ depends on the specific
realization of the alloy. We may take as a simple example
\begin{equation}
\nu(l) = \frac{V}{2 {\rm arctan}(\tilde{l}) } \frac{L}{l^2 + L^2} 
\label{nu-ansatz} 
\end{equation}
for $ |l| < \tilde{l} $ and $ \nu(l)=0 $ otherwise and $ - \tilde{l} <
l_0 < 0 $.
Eq.(\ref{equilibrium}) can solved numerically. For a given set of
parameters we display the results in Fig. \ref{phen-model-1}. 
In  Fig. \ref{phen-model-1}a) the volume of the
  F-domains as a function of $ H_{ex} $ is shown to grow monotonously
  and in  Fig. \ref{phen-model-1}b) the magnetization exhibits a
    clear metamagnetic transition at the point of fastest change of
    the volume. 

\subsection{Longitudinal Magnetoresistance} 

We now would like to consider the effect of domains on the transport 
in the metallic $\gamma $-band which we had
neglected so far. Naturally the conductance of the $ \gamma $-band
depends also on
the correlation of the localized degrees of freedom of the $
\alpha$-$\beta$-orbitals, since they 
interact via onsite couplings, such as Hund's rule coupling.
First of all the two uniform phases A and F would have slightly
different electrical conductance, $ \sigma_A $ and $ \sigma_F $. 
For the inhomogeneous phase the domain boundaries also play a role. 
They are restricted regions where orbital and
spin degrees of freedom have a rapid spatial dependence connecting the
two kinds of domains, so that the scattering of
electrons of the $ \gamma $-band in the domain boundary region is
enhanced. This readily understood from the point of view that the A-
and F-domains in a magnetic field have different magnetizations. Thus
the rapid change of magnetization in the domain boundary constitutes
a scattering potential.  

\begin{figure}
\resizebox{0.4\textwidth}{!}{%
  \includegraphics{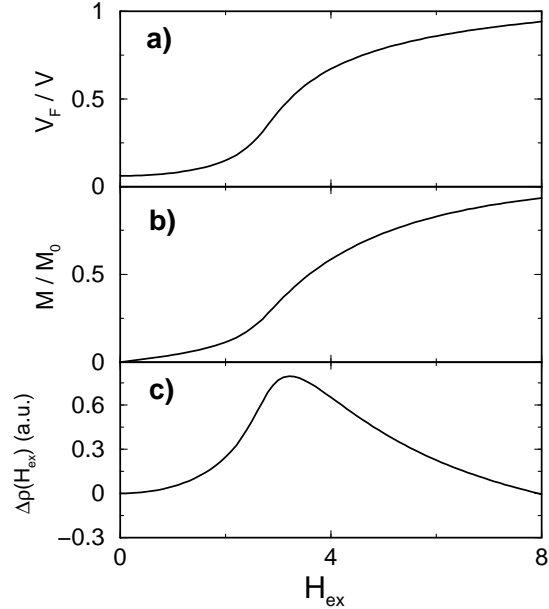}}
\caption{Results of the phenomenological model: a) Volume of the
  F-domain; b) magnetization ; c) longitudinal
  magnetoresistance. Parameters: $ W = 0.1 V $, $ L=2 $, $ l_0 = 2 $,
  $ \tilde{l}=2.5 $, $ M_{0F}=M_{0A} = 1 $, $ \chi_{OF}=0.3 $ and $
  \chi_{0A}= 0.02 $.}
\label{phen-model-1}
\end{figure}

We would like to consider the problem of the
longitudinal magnetoresistance, i.e. the resistance for a current
which flows parallel to the applied external (inplane) field. The change of the
domain distribution is the key ingredient determining the field
dependence of the resistance. We
address this problem again from the point of view of our simple
phenomenological model. The overall conductance of the sample
neglecting the domain boundaries may be roughly approximated by the
averaged conductance

\begin{equation}
\bar{\sigma} = \frac{V_F \sigma_F + (V-V_F) \sigma_A}{V}.
\end{equation}
if the conductances $ \sigma_F $ and $ \sigma_F $ have comparable
magnitude. Next we introduce the effect of the domain boundaries.
As additional scattering regions their influence depends on their
density. 
Hence, the resistance contains two parts coming from the domains and
the domain boundaries: 
\begin{equation}
\rho = \frac{1}{\bar{\sigma}} + \tilde{\rho}.
\end{equation}
Here $ \tilde{\rho} $ denotes the domain boundary part which
we approximate as proportional to the density of boundaries,
corresponding to $ \nu $: $ \tilde{\rho} = \rho_0 \nu $. Now we can discuss 
the field dependence of $ \rho $ by inserting the solution of
Eq.(\ref{equilibrium}), 
$ l(H_{ex}) $. This leads to a sensible
qualitative behavior of the longitudinal magnetoresistance as can be
seen in Fig.\ref{phen-model-1}c) showing 
$ \Delta \rho (H_{ex}) = \rho(H_{ex})- \rho(0) $. There is
a pronounced maximum at essentially the same position as the metamagnetic
transition occurs, both connected with the domain boundary percolation
condition.  
It is also worth noting that the low-field limit leads to

\begin{equation}
\Delta \rho (H_{ex}) \propto l-l_0 \propto  f_F(H_{ex}) - f_A (H_{ex})
\propto H_{ex}^2.
\end{equation}
This low-field behavior and the overall field dependence 
compares well with the experimental data by NM.\cite{nakatsu,naka-phd}
 In the experiment a pronounced maximum is
found at approximately the field corresponding to the metamagnetic
transition. This strong feature in the magnetoresistance disappears
for temperature higher than the characteristic temperature of the
antiferromagnetic spin correlation ($ \sim 10 K $). Naturally the
domain formation is absent when there is no spin correlation (AFM
order in our mean field treatment).

\subsection{Resistor network approach}

The issue of transport in such a system is naturally complex since it
involves percolation of domains and domain boundaries.
This aspect is not taken into account in our
model where we 
the domain boundaries only considered as additional uncorrelated scatterers 
entering the resistance only through their density.
The domain boundary, however, should 
be considered as a barrier the electron would have to traverse. 
Hence, it would redirect the current
flow which is difficult to include in our simple model. 
In order to demonstrate, however, that the basic properties of the electrical 
resistance are not changed by including this aspect, 
we consider a two-dimensional resistor network model where we can 
implement the progressive change of the conductivity and the 
presence of domain boundaries as barriers in a simple way.\cite{BURGY}

\begin{figure}
\resizebox{0.4\textwidth}{!}{%
  \includegraphics{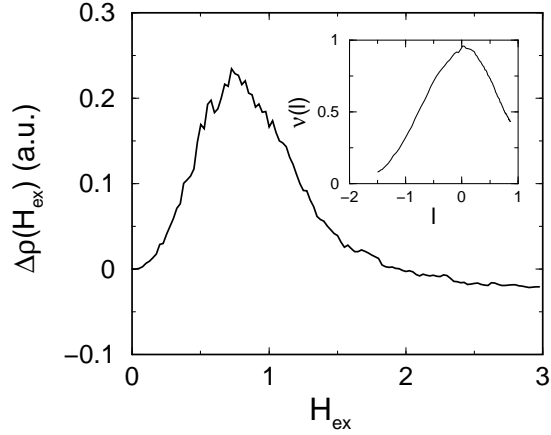}}
\caption{Magnetoresistance for a random sample of the size $ N \times
  L = 100 \times 100 $ ($ R_A : R_F : R_{db} = 1 : \frac{4}{3} : 20
  $). Inset: $ \nu(l) $ defined as the density of 
  domain boundaries is here proportional to the number of domain
  boundary resistors in the system for a given domain formation.}
\label{resnet}
\end{figure}

The network is a system of $ N \times L $ knots on square lattice
which coupled to their four nearest neighbors via resistors. The knots 
lie either in the A- or F-domain. Two neighboring knots in the same
domain are connected via a resistor with the corresponding domain
resistance $ R_A $ or $ R_F $, respectively, while the resistor
between neighboring 
knots of different domains has a resistance  $ R_{db} $ which is considerably
larger, corresponding 
to the barrier effect of the domain boundary. We assume a potential
difference along $ L $-direction of the network and apply
Kirchhof's law which yields a linear current voltage
relation, i.e. the resistance. 
The domain distribution is generated by means of a real function $ Q(n_x ,
n_y) $ where $ n_x $ and $ n_y $ are integers denoting the
knots. Introducing 

\begin{equation}
\tilde{Q}(n_x , n_y; l) = Q(n_x ,n_y) + l
\end{equation}
we define a site $ (n_x,n_y) $ to belong to the A(F)-domain, if the 
$ \tilde{Q}(n_x,n_y; l) $ is positive (negative). The parameter $ l $
grows as a function of the field and so changing the domain
distribution, in the very same way as we have seen in our very simple
model above. Obviously, defining a $ \tilde{\nu}(l) $ here as the
number of bonds in the network with a resistor $ R_{db} $ yields
a function equivalent to $ \nu(l) $ assumed above (see inset of
Fig.\ref{resnet}). 

The function $ Q(n_x , n_y ) $ is generated from random numbers by 
smoothening in the same way as the strain distribution for our
mean field discussion. We show the results for a simulation for $ N
\times L = 100 \times 100 $ and take 

\begin{equation}
l - l_0 = \frac{aH_{ex}^2}{1+b H_{ex}^2}
\end{equation}
in order to mimic the field dependence of the parameter $ l $ which
should have a qualitatively similar behavior as the $ l $ in
Eq.(\ref{equilibrium}), i.e.  
$ l - l_0 $ is quadratic in the field for small fields and grows
faster around the percolation regime ($ H_{ex} $ dimensionless and $
a=1/4 $ and $ b=1/2 $). 
In Fig.\ref{resnet} we show the result for the resistivity with the parameters
given in the caption. The same qualitative behavior is
found here as in Fig. \ref{phen-model-1} including the behavior of $
\nu(l) $, demonstrating that the simple phenomenological model
captures the basic properties of the role of domain boundaries in the
transport.

\section{Discussion of the experiment}

Our model reveals two essential points which we can compare with the
experimental situation of $ Ca_{2-x} Sr_x  RuO_4 $. (1) We obtained a
mean field phase diagram as a function of temperature and orthorhombic
deformation which we can translate to the experimental phase diagram of
temperature versus Sr-concentration $x$ for $ 0.2 \leq x \leq 0.5 $. (2)
We identify our antiferromagnetic phase (with driven ferro-orbital
correlation) with the region II of the
experimental phase diagram and can discuss some important
properties. For both points the disorder effect of the alloy plays a
decisive part.  

(1) The concentration $ x = 0.5 $ is separating region III and region
    II. At this point the system is still dominantly tetragonal and
    the domains with anti-ferro-orbital order dominate. The spin
    susceptibility is Curie-like and shows at temperatures $ \sim 1 K
    $ a irreversible behavior. This low-temperature phase  
    was interpreted by Nakatsuji et
    al. as a cluster glass. Within our model we would rather identify
    this  phase with an inhomogeneous ferromagnetic
    phase. The observed slow dynamics is then not a glass-like
    feature, but rather an effect due to slow motion of weakly pinned
    domain walls. 
    If the concentration $ x $ decreases into the region II, gradually
    the size of the antiferro-orbital domains is shrinking and the
    ferromagnetic domains eventually do not percolate anymore, so that the
    irreversible low-temperature phase ceases to exist. 

(2) Close to $ x =0.2 $ in region II the physics is governed by the
    dominance of a percolating domain with antiferromagnetic spin
    and the driven ferro-orbital correlation. Domains of
    antiferro-orbital order and the ferromagnetic spin correlation are
    sparse. In this situation we find two important features to
    compare with experiments. The first is the reduced crystal
    symmetry which due to spin-orbit coupling manifests itself in the
    anisotropy of the spin susceptibility. (A puzzle remains in the fact
    that the orthorhombic distortion is untwinned in all the samples
    investigated so far, so that this anisotropy is very pronounced.\cite{NAKAMAE,FRIEDT2})
    The second feature is connected with the
    fact that the size of the minority domains can be increased by
    an external magnetic field. As a consequence a
    metamagnetic transition is observed at roughly $2.5 T $ for 
    fields along the inplane direction with the maximal spin
    susceptibility and a pronounced peak in the longitudinal
    magnetoresistance around $ 2.7 T $. In our previous discussion we
    have shown that these features fit well into the picture of
    redistributing domains by the magnetic field. In particular, 
    these effects disappear for temperatures exceeding $10 K $ which
    corresponds to the maximum of the uniform spin susceptibility
    indicating the onset of antiferromagnetic correlations. Only
    below this characteristic temperature domain formation is possible
    at all. 

Finally, we would like to comment on the energy scales entering our
effective Hamiltonian. Our mean field treatment yields real phase
transitions whose onset temperatures can be taken as the
energy/temperature scale. Identifying the Ne\'el temperature as
approximately $ 10 K $ we may choose $ J $ to be about 
$ 30 K $.  Then, the onset of FM order around $x=0.5$ is of
the order of $ 1 K $  and the critical temperature of the AFO order
lies between $ 10 $ and $ 20 K $.  
While the magnetic orders are clearly related to the experimentally
observed behavior of the 
system, the staggered orbital order has not been observed so
far. Nevertheless, it is important to remark here that the LDA+U 
calculation  by Anisimov et al. based on the experimentally determined
crystal structure data at $ x=0.5 $,  
clearly reveals the staggered orbital correlation as
a dominant orbital feature including simultaneously ferromagnetism.\cite{anis2}

\section{Conclusions}

The initial motivation to explain the presence of spin 1/2 observed in
the Curie-like 
susceptibility at the boundary between region II and III, has led to
the hypothesis of a Mott insulator which is restricted to two of
three orbitals. This would introduce a localized spin 1/2 and orbital
degrees of freedom besides a band of itinerant electrons. We have 
seen that the resulting effective model (Eq.(\ref{Heff})) does not
only explain the Curie-like susceptibility 
around $ x = 0.5 $, but gives a surprisingly good account of a variety
of additional features: a qualitative description of the NM phase diagram of
Ca$_{2-x}$Sr$_x$RuO$_4$ as well as a good account of the
properties of region II in a magnetic field.  
We do not claim that the features discussed successfully by this model
is a proof of validity of the initial hypothesis of orbital
selective Mott transition. It is definitely also 
important to take in future studies the
coupling between the itinerant $ \gamma $-band and the localized
degrees of freedom, as it would influence the phase diagram obtained. 
Nevertheless, we believe that our discussion provides some evidence
that localized degrees of freedom should be involved in the physics
of this material. It seems difficult to us to
explain these characteristic properties of Ca$_{2-x}$Sr$_x$RuO$_4$ by
a model entirely based on itinerant electrons.

We are very grateful to S. Nakatsuji, Y. Maeno, M. Braden, O. Friedt
and L. Balicas for many helpful discussions. Especially we thank
T.M. Rice, V.I. Anisimov, I.A. Nekrasov and D.E. Kondakov for many
illuminating discussion during the setup of this work. M.T. acknowledges the Aspen Center for Physics and the KITP where parts of this work have been carried out. 
We would also like to thank for   
financial support by the Swiss Nationalfonds and a grant-in-aid of the
Japanese Ministry of Education, Science, Culture and Sports.

\appendix
\section{Strain and isospin}

We consider here the symmetry properties of the terms containing
linear couplings of the isospin to the strain and the magnetic field
and spin. It is important to notice that ( $ | + \rangle , | - \rangle
$) transform under symmetry operations of the RuO$_2$-plane ($C_{4v}
$) like the coordinates $ (x,y)$. Therefore is the basic symmetry
property of the isospin components: 

\begin{equation} \begin{array}{ll}
I^z = \frac{1}{2} ( |+ \rangle \langle + | -  | - \rangle \langle - |
) & \to x^2 - y^2 \\ & \\
I^x = | + \rangle \langle - | + | - \rangle \langle + | &  \to
xy
\end{array}
\end{equation}
We ignore here the $y$-component of the isospin, since we will not use
it. From this follows that the allowed couplings to the invariant
terms in the coupling to the strain result from $ \epsilon_1 =
\epsilon_{xx} - \epsilon_{yy} $ for $ I^z $ and $ \epsilon_2 =
\epsilon_{xy} $ for $ I^x $. Thus the coupling term has the general
form

\begin{equation}
K_1 I^z \epsilon_1 + K_2 I^x \epsilon_2 
\end{equation}
with $ K_{1,2} $ as phenomenological coupling constants. The coupling
between strain and isospin happens via the crystal field level
splitting. 

\section{Spin-orbit coupling}

By the same symmetry argument we can derive the allowed terms
describing the effect of spin-orbit coupling as anisotropic coupling
of the magnetic field to the spin,

\begin{equation} \begin{array}{l}
I^z \qquad \leftrightarrow \qquad H_x S_x - H_y S_y \\ \\
I^x \qquad \leftrightarrow \qquad H_x S_y + H_y S_x
\end{array}
\end{equation}
which leads to

\begin{equation}
g_1 \langle I^z \rangle (H_x S_x - H_y S_y) + g_2 \langle I^x \rangle
(H_x S_y + H_y S_x) 
\end{equation}
where again $ g_{1,2} $ are phenomenological constants. Note that the 
effect of spin-orbit coupling also involves the itinerant $ \gamma
$-orbital. This is automatically taken care in this symmetry
consideration.

\end{document}